\documentclass[aps,twocolumn,pra,tightenlines,floatfix]{revtex4}
\usepackage[dvips]{graphicx}
\usepackage[english]{babel}
\usepackage{amsmath}
\usepackage{amssymb}
\usepackage{times}
\newcommand{\bs}{\begin{split}}
\newcommand{\es}{\end{split}}
\newcommand{\bea}{\begin{eqnarray}}
\newcommand{\eea}{\end{eqnarray}}
\newcommand{\be}{\begin{equation}}
\newcommand{\ee}{\end{equation}}
\newcommand{\ba}{\begin{eqnarray}}
\newcommand{\ea}{\end{eqnarray}}
\newcommand{\ek}{\epsilon_{\mathbf{k}}}

\newcommand{\uk}{u_{\mathbf{k}}}
\newcommand{\vk}{v_{\mathbf{k}}}
\newcommand{\sumk}{\sum_{\mathbf{k}}}

\newcommand{\p}{\partial}
\newcommand{\D}{\mathrm{D}}

\newcommand{\doublebar}[1]{\bar{\bar{#1}}}

\newcommand{\Ekq}{E_{\mathbf{kq}}}

\newcommand{\xikq}{\xi_{\mathbf{kq}}}
\newcommand{\xik}{\xi_{\mathbf{k}}}

\begin{document}

\title{Single-plane-wave Larkin-Ovchinnikov-Fulde-Ferrell state in
  BCS--Bose-Einstein condensation crossover}

\author{Yan He, Chih-Chun Chien, Qijin Chen and K. Levin}

\affiliation{James Franck Institute and Department of Physics,
 University of Chicago, Chicago, Illinois 60637}

 \date{\today}

\begin{abstract}
  We study the single-plane-wave Larkin-Ovchinnikov-Fulde-Ferrell (LOFF)
  states for BCS--Bose-Einstein condensation (BEC) crossover at general
  temperatures $T$.  Because we include the important effects of
  noncondensed pairs, our $T \neq 0$ phase diagrams are different from
  those reported in earlier work.  We find that generalized LOFF phases
  may be the ground state for a wide range of (weak through moderately
  strong) interactions, including the unitary regime. However, these
  LOFF phases are readily destroyed by non-zero $T$.
\end{abstract}

\pacs{03.75.Hh, 03.75.Ss, 74.20.-z \hfill \textsf{\textbf{cond-mat/0610274}}}

\maketitle

Recent atomic physics discoveries in the field of ultracold polarized
\cite{ZSSK06,PLKLH06,ZSSK206} fermionic superfluids have important
implications in color superconducting quark matter as well as in dense
nuclear matter \cite{LW03,FGLW05,LOFF_Review}.  Moreover, there has been
a long standing interest from the condensed matter community
\cite{Combescot} in observing the very elusive
Larkin-Ovchinnikov-Fulde-Ferrell (LOFF) \cite{FFLO} states of a
polarized superfluid. Here condensation of Cooper pairs takes place at
one or more non-zero momenta $\mathbf{q}_i $.  These cold gases possess
a remarkable flexibility in which they can be polarized as well as
studied with variable attractive interaction. This provides an
additional mechanism for (possibly) tuning in the various LOFF phases as
one changes the $s$-wave two-body scattering length $a$ from positive in
the Bose-Einstein condensation (BEC) regime to negative (BCS).  Thus
far, experiments \cite{ZSSK06,PLKLH06,ZSSK206} have focused on the
unitary scattering regime, midway from BCS to BEC.  While Bogoliubov-de
Gennes (BdG) based theories \cite{Machida2,Kinnunen} for a trap at
unitarity suggest that the ground state is generalized LOFF, homogeneous
studies \cite{SR06} conclude that LOFF1 at least, is confined to a
sliver near the BCS endpoint.  A critical component which needs to be
injected into this controversy is the nature of the stability criteria
\cite{Pao,Gubankova,us,Radstability} which is, similarly, under lively
debate.

The goal of this paper is to clarify these issues by addressing BCS-BEC
crossover at general $T$. We focus on one particular member of the LOFF
class-- corresponding to the single plane wave LOFF state, hereafter
called LOFF1, in a homogeneous system.  We characterize the ``existence
regime" (where solutions exist) and the ``stability regime" (where
solutions are stable) in a series of phase diagrams. Unlike Ref.
\cite{Radlong} our calculations do not automatically incorporate first
order transitions (from a single phase state) to a phase-separated
state.  This phase separation is not relevant to all sub-disciplines
\cite{LOFF_Review} and in future applications to a trap, one has also to
include surface energy terms which are difficult to estimate.  Moreover,
phase separation has not been included in the Bogoliubov deGennes (BdG)
based theories \cite{Machida2,Kinnunen} which we want to understand
here.

However, with phase separation included, our $T=0$ results are rather
similar to those found in Ref.~\cite{Radlong}. Here, we emphasize finite
$T$, and we implement a highly numerical procedure to solve all coupled
equations directly at fixed total particle number $N_\sigma$.  Even in
the absence of alternative phase separation states, the \emph{stable}
LOFF1 state is primarily restricted to a regime near the BCS endpoint,
although it does overlap unitarity for a narrow range of high
polarizations. We show that the LOFF1 existence regime is considerably
broader and is directly associated \cite{Tsinghuagroup} with the phase
space region where there is negative superfluid density
\cite{Pao,Chien_prl} in the $\mathbf{q}=0$ or Sarma state
\cite{Sarma63}.  Here it is likely that a LOFF phase of one form or
another will be stable, although it may be something more complex
\cite{Combescot,LOFFlong} than LOFF1.

Our central phase diagram in the $T$ vs $p$ plane should be of
particular interest to experimentalists who are currently creating plots
of this nature.  It can be contrasted with that obtained in Ref.
\cite{Parish} in which temperature was introduced in a fashion following
the original Nozieres--Schmitt-Rink (NSR) \cite{NSR} scheme. Here,
unlike Ref.~\cite{Parish} we choose to include $T$ in a manner
which is fully consistent with the very extensive 
literature 
\cite{Pao,SR06,Kinnunen,Machida2}
on the ground state of these polarized superfluids.

We introduce $ T \neq 0$ following a $T$-matrix scheme, and restrict our
attention to the superfluid phase.  This $T$-matrix represents the
propagator for non-condensed pairs and is given by
$t^{-1}(P)=U^{-1}+\chi(P) $ where $\chi$ is the pair susceptibility and
$U<0$ is the pairing interaction strength. For atomic Fermi gases, we
assume an $s$-wave contact interaction.
There are strong similarities between $t(P)$ and the Hartree-Fock
approximation to the particle-hole susceptibility which involves the
usual Lindhard function and on-site repulsion. These Hartree-Fock
theories are used to establish whether ferro- or antiferromagnetic order
will arise.  Here we consider a very similar competition between Sarma
and LOFF1 states. The relevant $\chi(P)$ necessarily involves the self
consistently determined fermionic gap parameter $\Delta(T)$ and chemical
potential $\mu(T)$.  Importantly at and below $T_c$ the chemical
potential for the \textit{pairs} ($\mu_{pair}$) must be zero and this
BEC condition on $t(P)$, thereby, determines $\Delta(T)$.

The pair susceptibility for LOFF1 condensates in which momentum
$\mathbf{k}$ pairs with $\mathbf{-k+q}$, (for, as yet undetermined
$\mathbf{q}$) may readily be written down \cite{LOFFlong}.  We first
introduce the fermionic chemical potentials $\mu_\uparrow$ and
$\mu_{\downarrow}$ for the two spin states, 
$\mu=(\mu_{\uparrow}+\mu_{\downarrow})/2$ and $h=(\mu_{\uparrow}
-\mu_{\downarrow})/2$,
and $\ek=\mathbf{k}^2/2m$, $\xik=\ek-\mu$, where $\mu_\sigma$ is the
chemcal potential for spin $\sigma=\uparrow,\downarrow$. It is useful to
also define $\Ekq=\sqrt{\xikq^{2}+\Delta^{2}}$, with
$\xikq=(\xik+\xi_\mathbf{k-q})/2$ and $\delta\ek = (\ek-\epsilon_{\bf
  k-q})/2$.  As in Ref.~\cite{Chien06}, we set the volume $V=1$,
$\hbar=k_B = 1$, and $P\equiv (i\Omega_l, \mathbf{p})$, where
$\Omega_l=2l\pi T$ is an even Matsubara frequency.
Then the pair susceptibility $\chi(P)$ at the mean field level, after
analytical continuation $i\Omega_l \rightarrow \Omega + i 0^+$, is given
by
\begin{eqnarray}\label{eq:LOFF1chi}
  \chi(P)&=&\sumk\left[
    \uk^{2}\frac{\bar{f}(E_{kq}+\delta\ek)+\bar{f}(\xi_{\mathbf{p-k}})-1}
    {\Omega-\xi_{\mathbf{p-k}}-(\Ekq+\delta\ek)} \right.
  \nonumber\\ &&\left.{}
    +\vk^{2}\frac{\bar{f}(\xi_{\mathbf{p-k}})-\bar{f}(\Ekq-\delta\ek)} 
    {\Omega-\xi_{\mathbf{p-k}}+(\Ekq-\delta\ek)}\right],
\end{eqnarray}
which, as $\mathbf{q} \rightarrow 0$, goes over smoothly to its
counterpart in the Sarma phase.  Here the coherence factors $\uk^{2},
\vk^{2}=(1\pm \xikq/\Ekq )/2$ and we define $ \bar{f}(x) =
[f(x-h)+f(x+h)]/2$, where $f(x)$ is the Fermi distribution function.
The BEC condition, $U^{-1}+\chi(0,\mathbf{q})=0$,
leads to the gap equation which, when written in terms of the scattering
length $a$, is of the standard form in the literature
\begin{equation}\label{eq:gap}
  -\frac{m}{4\pi a}=\sumk \left[\frac{1-2\doublebar{f}(\Ekq)}{2\Ekq}
    -\frac{1}{2\epsilon_{k}}
  \right] \,,
\end{equation}
where we define $ \doublebar{f}(x) \equiv [f(x-h+\delta\ek ) +
f(x+h-\delta\ek)]/2 $.

The number equations also depend on the quantity $\chi$ through a self
energy involving $t(P)$; we show the details elsewhere \cite{LOFFlong}
but summarize our final results which yield the standard equations in
the literature
%\begin{subequations}\label{eq:num}
\begin{eqnarray}\label{eq:num}
  n &=& 2\sum_\mathbf{k} \left[\vk^2 + \frac{\xikq}{\Ekq}
    \doublebar{f}(\Ekq)\right],\\ 
  \delta n &=& \sumk\delta f(\Ekq)\,,\label{eq:ndiff}
\end{eqnarray}
%\end{subequations}
%
where $n=n_\uparrow + n_\downarrow$ and $\delta n=n_\uparrow -
n_\downarrow$, and the polarization $p\equiv \delta n/n$.  Here we have
defined
$  \delta f(x) = [f(x-h+\delta\ek )-
  f(x+h-\delta\ek)]. $
Finally, we determine $\mathbf{q}$ by imposing
an extremal condition on the pair susceptibility,
$ \left.\frac{\partial\chi(0,\mathbf{p})} {\partial\mathbf{p}}
\right|_{\mathbf{p=q}}=0. $ This condition, which turns out to be
equivalent to requiring that there be no net current in this LOFF1
state,
is given by 
\begin{eqnarray}\label{eq:mom}
  0&=&\frac{1}{\Delta^2}\sumk\bigg\{\frac{\mathbf{q}}{2}\left[
  \Big(1-\frac{\xikq}{\Ekq}\Big)-
  \frac{\xikq}{\Ekq}\doublebar{f}(\Ekq)\right] \nonumber\\  
  &&{}+\left(\mathbf{k-}\frac{\mathbf{q}}{2}\right)\delta f(\Ekq)\bigg\}\,.
\end{eqnarray}
When this equation has a solution at $\mathbf{q} \neq 0$ we have a LOFF1
phase.  There will always be a co-existing solution of the Sarma-type
with $\mathbf{q} =0$.

\begin{figure}
  \centerline{\includegraphics[width=2.9in,clip]{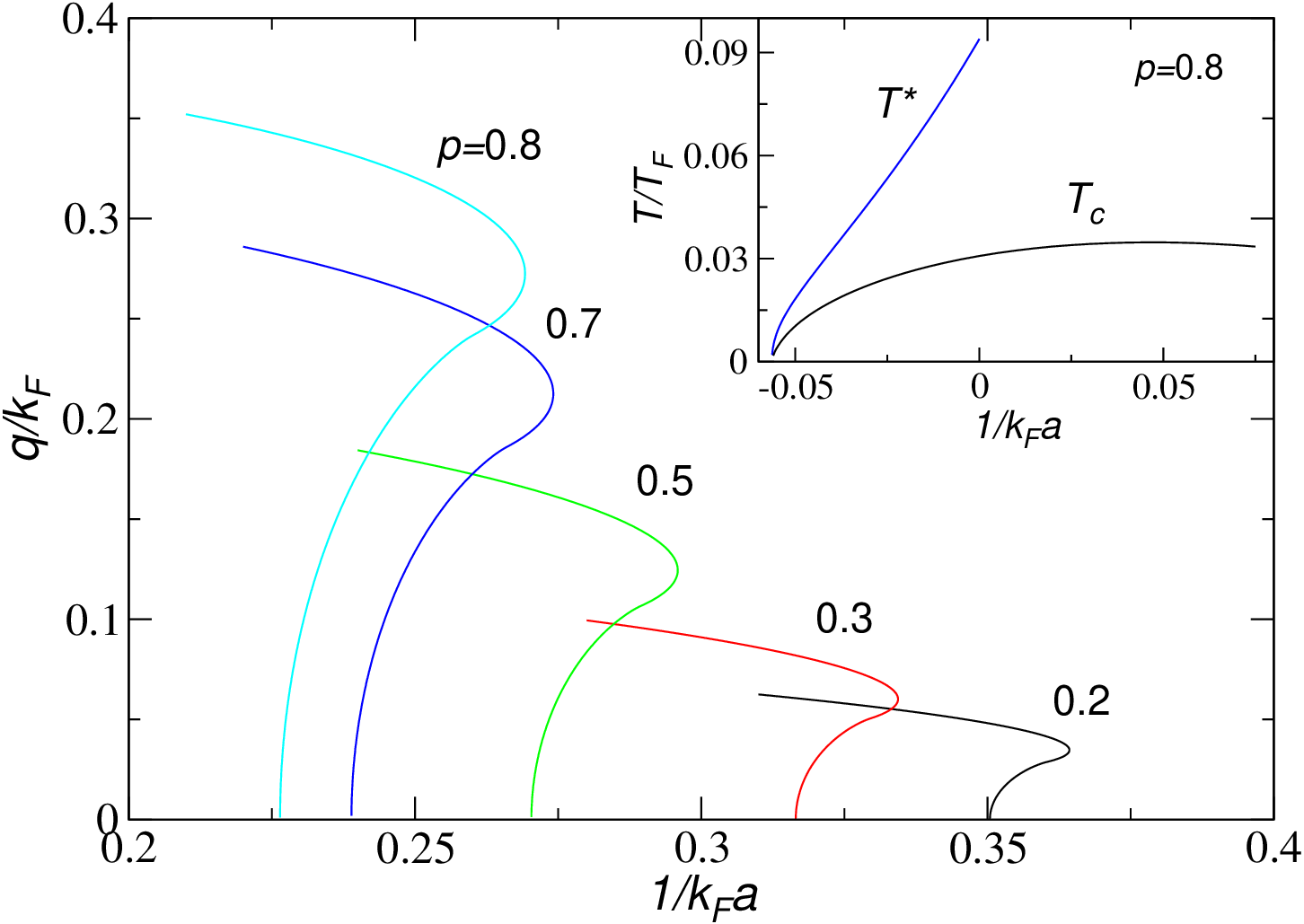}}
  \caption{LOFF1 wavevector $q$ as a function of $1/k_Fa$ at $T=0$ and
    for various $p$. Beyond the turning point, no LOFF1 state exists.
    The inset shows $T_c$ and mean field $T_c^{MF} =T^*$ (pair formation
    temperature) at unitarity over respective stability regimes.}
\label{fig:qdelta}
\end{figure}

At $T \neq 0$, the parameter $\Delta(T)$ contains the contribution from
both condensed (sc) and non-condensed (pg) pairs. We can show quite
generally that below $T_c$, $\Delta^2 (T) \equiv \Delta_{sc}^2(T) +
\Delta_{pg}^2(T)$, where
\begin{equation}\label{eq:pg}
  \Delta_{pg}^2 (T) = Z^{-1}\sum_{\mathbf{p}} b(\Omega_{\bf p}) \,,
\end{equation}
and $b(x)$ is the Bose distribution function.  Here the pair dispersion
is found to be $\Omega_{\mathbf{p}} \approx ({\bf p-q})^2/2 M^*$.  Analytical
expressions for $M^*$ and $Z$ are possible via an 
expansion of $\chi$ in small $({\bf p-q})$,
\begin{eqnarray}
  \chi(\Omega, \mathbf{p})-\chi(0,\mathbf{q}) &\approx& Z \left[\Omega  - \frac{(\mathbf{p-q})^2}{2M^*}\right],
%g^{-1}_{eff}(P) - g^{-1}_{eff}(0) &=& Z_g (i\Omega_n  -B_g q^2) \nonumber
\end{eqnarray}
where $Z=\left.\frac{\partial
    \chi}{\partial\Omega}\right|_{\Omega=0,\mathbf{p=q}}$ and
$\frac{1}{2M^*} =-\left.\frac{1}{6Z}\frac{\partial^{2} \chi}{\partial
    \mathbf{p}^{2}}\right|_{\Omega=0,\mathbf{p=q}} $.
The quantity $\chi$ contains everything one needs to know about zero as
well as finite $T$.  And our results for $T=0$ reduce to the standard
equations in the literature.

\begin{figure*}
  \centerline{\includegraphics[width=6.in,clip]{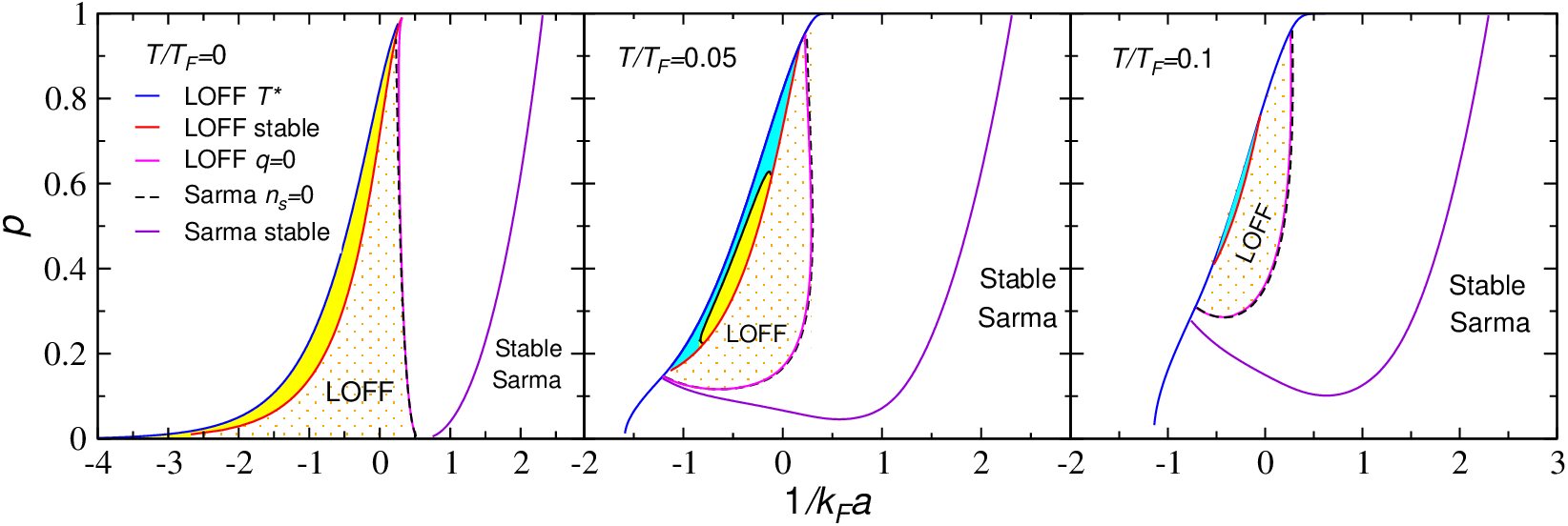}}
  \caption{(Color online) Phase diagram in the $p$ vs $1/k_Fa$ plane for
    $T/T_F=0$, 0.05, and 0.1. The dotted region shows where LOFF1 state
    is unstable but some form of LOFF phase may be stable. The (yellow)
    light shaded region indicates the stable LOFF1 superfluid and the
    (cyan) darker shaded region the stable LOFF1 normal state. 
    }
\label{fig:Tphase}
\end{figure*}

To demonstrate that a given LOFF1 solution to our self consistent
equations is stable, we introduce an effective
thermodynamic potential for this superfluid state
\begin{eqnarray}\label{eq:thermo}
   \Omega&=&-\frac{\Delta^2}{U\;}+\sumk\big\{(\xikq-\Ekq)\nonumber\\
  & &{}-T\ln (1+\exp[-(\Ekq-h+\delta\ek)/T])\nonumber\\
  & &{}-T\ln (1+\exp[-(\Ekq+h-\delta\ek)/T])\big\}\,.
%& &+T\ln (1+e^{-\xi_{\mathbf{k}\uparrow}/T})+T
%\ln (1+e^{-\xi_{\mathbf{k}\downarrow}/T})]
\end{eqnarray}
It is straightforward to verify that the above gap, number and
zero-current equations are consistent with the variational conditions
$$\frac{\partial\Omega}{\partial\Delta}=0;\mbox{ }\mbox{
}-\frac{\partial\Omega}{\partial\mu}=n; \mbox{ }\mbox{
}-\frac{\partial\Omega}{\partial h}=\delta n; \mbox{ }\mbox{
}\frac{\partial\Omega}{\partial \mathbf{q}}=0.$$
The stability condition
requires that the symmetric number susceptibility matrix
\begin{equation}\label{numsus}
M=\begin{pmatrix} \frac{\D N}{\D
\mu} & \frac{\D N}{\D h} \\ \\ \frac{\D \delta
N}{\D \mu} & \frac{\D \delta N}{\D h}
\end{pmatrix} 
\end{equation}
be positive definite \cite{Pao,Gubankova}.
Here $\frac{D\;}{\D x} \equiv \frac{\p\; }{\p x} + \frac{\p\Delta }{\p
  x} \frac{\p\; }{\p \Delta} +  \frac{\p \mathbf{q} } {\p
  x} \cdot\frac{\p\; }{\p \mathbf{q}} $, with $x = \mu, h$.
To evaluate this matrix we note
%\pagebreak
%\begin{subequations}
\begin{eqnarray}
  \frac{\D N}{\D\mu}&=&
%\frac{\p N}{\p\mu}+\frac{\p N}{\p\Delta}
%\frac{\p\Delta}{\p\mu}+\frac{\p N}{\p q}\frac{\p q}{\p\mu}
  -\frac{\p^2 \Omega}{\p\mu^2}-\frac{\p^2\Omega}{\p\Delta\p\mu}
  \frac{\p\Delta}{\p\mu}
  -\frac{\p^2\Omega}{\p q\p\mu}\frac{\p q}{\p\mu},\nonumber\\
%\end{eqnarray}
%\begin{eqnarray}
  \frac{\D N}{\D h}&=&
%\frac{\p N}{\p h}+\frac{\p N}{\p\Delta}
%\frac{\p\Delta}{\p h}+\frac{\p N}{\p q}\frac{\p q}{\p h}
  -\frac{\p^2 \Omega}{\p\mu\p h}-\frac{\p^2\Omega}
  {\p\Delta\p\mu}\frac{\p\Delta}{\p h}
  -\frac{\p^2\Omega}{\p q\p\mu}\frac{\p q}{\p h} = \frac{\D\delta
    N}{\D\mu},\nonumber \\
%\end{eqnarray}
%\begin{eqnarray}
  \frac{\D\delta N}{\D h}&=&
%\frac{\p\delta N}{\p h}
%+\frac{\p\delta N}{\p\Delta}\frac{\p\Delta}{\p h}
%+\frac{\p\delta N}{\p q}\frac{\p q}{\p h}
  -\frac{\p^2 \Omega}{\p h^2}-\frac{\p^2\Omega}{\p\Delta\p h}
  \frac{\p\Delta}{\p h}
  -\frac{\p^2\Omega}{\p q\p h}\frac{\p q}{\p h},\nonumber
\end{eqnarray}
%\end{subequations}
%
where $\p\Delta/\p\mu$, $\p q/\p \mu$, $\p\Delta/\p h$, and $\p q/\p h$
can be easily derived by differentiating Eqs.~(\ref{eq:gap}) and
(\ref{eq:mom}) with respect to $\mu$ and $h$. It can be shown that the
positive definiteness of $M$ is equivalent to
\begin{equation} \label{eq:11}
\frac{\p^2\Omega}{\p\Delta^2}\frac{\p^2\Omega}{\p q^2}
-\Big(\frac{\p^2\Omega}{\p\Delta\p q}\Big)^2 >0 \,,
\end{equation}

Equations (\ref{eq:gap})-(\ref{eq:pg}) were solved numerically to
establish the existence regime for LOFF1 solutions.  Figure
\ref{fig:qdelta} shows a plot of the behavior of the LOFF1 wavevector
$\mathbf{q}$ as a function of $1/k_Fa$, and at $T=0$, for various
polarizations $p$. Here $E_F=k_BT_F=\hbar^2k_F^2/2m$ is defined as the
Fermi energy of an unpolarized, noninteracting Fermi gas of density $n$.
For each value of $p$ a turning point, $(1/k_Fa)_{max}$, is visible
beyond which we can not find LOFF1 solutions.  At finite $T$ the
analogous curves (not shown) quickly become monotonically decreasing so
that $(1/k_Fa)_{max}$ corresponds to $\mathbf{q}=0$; thus the system
smoothly transforms to the Sarma state.

The inset in Fig.~\ref{fig:qdelta} shows the behavior of $T_c$ and its
mean field counterpart $T^*= T_c^{MF}$ as a function of $1/k_Fa$ near
unitarity and for $p=0.8$.  It can be seen that there is a considerable
difference between $T_c$ and $T_c^{MF}$ showing that pair fluctuation
effects (via $\Delta_{pg} \neq 0$) are very important in the LOFF1
phase, just as seen elsewhere \cite{Chien_prl}.  In contrast to the
behavior of the Sarma phase at unitarity, in the LOFF1 state,
superfluidity extends over the range of temperatures from $T_c$ down to
$T=0$.

\begin{figure*}
\centerline{\includegraphics[width=6.in,clip]{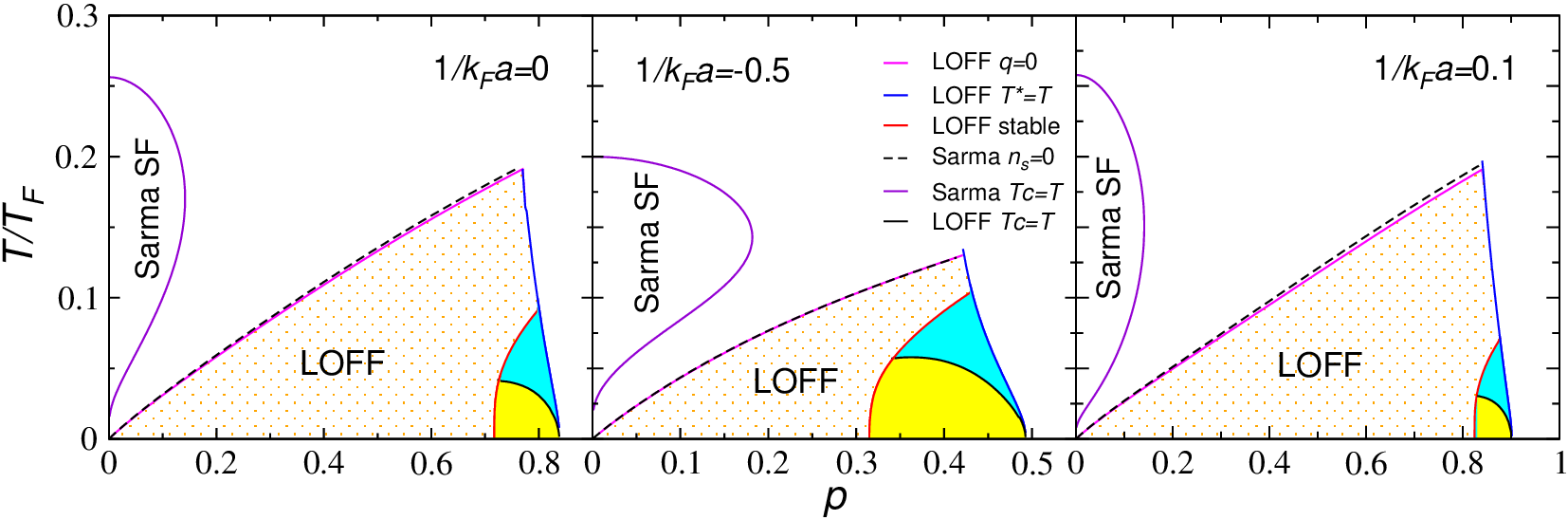}}
\caption{(Color online) Phase diagram in the $p$ -- $T$ plane for
  unitary (left), near BCS (middle) and near-BEC (right). The dotted
  region shows where LOFF1 state is unstable but stable generalized LOFF
  phases may in principle exist.  The (yellow) light shaded region is
  the LOFF1 superfluid, and the (cyan) darker shaded region is the LOFF1
  normal state.}
\label{fig:kaphase}
\end{figure*}

We turn to Fig.~\ref{fig:Tphase} beginning with the left panel ($T=0$)
and the nearly vertical line which is determined from $(1/k_Fa)_{max}$.
This provides a bound on the existence region for our numerically
obtained LOFF1 solutions. Essentially on top of this nearly vertical
line is the locus of points to the left of which the superfluid density
for the Sarma state $n_s^{Sarma}(0)$ is negative. That these two lines
coincide reinforces earlier work \cite{Tsinghuagroup}: providing one
considers a second order transition between generalized LOFF and Sarma
states, the boundary line for the existence of \textit{all} LOFF states
is determined by $n_s^{Sarma}=0$.  We indicate by the dotted background
in Fig.~\ref{fig:Tphase} where we have the possibility of superfluidity
with multiple nonzero $\mathbf{q}$'s.  It should be noted that this
existence regime for generalized LOFF states is relatively wide at
$T=0$, importantly including unitarity.
The shaded region in Fig.~\ref{fig:Tphase} results from applying the
stability criteria associated with the positivity of the matrix $M$ in
Eq.~(\ref{numsus}).  The non-vertical line to the right in this first
panel marks the onset point for a stable Sarma phase.

This phase diagram evolves with temperature as shown by the other two
panels in Fig.~\ref{fig:Tphase}.  In the middle and right panels we have
distinguished between normal and superfluid LOFF1 states by using darker
and lighter shaded regions respectively.  At the highest $T$, for the
panel on the right, there is only a sliver of stable LOFF1 which
corresponds to a normal (pseudogapped) phase.  It should be noted that
the size of the existence region for generalized LOFF solutions quickly
decreases as temperature is raised. This is related to the fact that the
$n_s^{Sarma}$ rapidly becomes non-negative as $T$ increases from zero.

Figure \ref{fig:kaphase} represents a particularly convenient way of
presenting our results. For the trapped case, experimental studies
\cite{ZSSK06,PLKLH06,ZSSK206} are in the process of mapping out this
phase diagram in the $p$--$T$ plane at unitarity.  From left to right,
the three panels correspond to unitarity, and the BCS and BEC sides
(close to resonance), respectively.  Also shown here is the region where
we have a stable Sarma state.  This appears only at intermediate
temperatures \cite{Chien_prl}, when the superfluid density (which is
negative at $T=0$ for all 3 cases) is driven positive.  Note that our
$p$--$T$ phase diagram is different from that in Ref.~\cite{Parish},
which is based on a different but unspecifed ground state.
Indicated in all three panels are the (dotted) regions where generalized
LOFF states may exist, and the (shaded) regimes where the LOFF1 phase is
stable.  As in Fig.~\ref{fig:Tphase}, the light shaded region
corresponds to the superfluid LOFF1 phase, and the dark shaded region to
the normal LOFF1 phase with a pseudogap. It is evident that, for these
$1/k_Fa$ values, a stable LOFF1 phase exists only at relatively high $p$
and low $T$. This is to be contrasted with the stable Sarma superfluid
which exists only at low $p$ and intermediate $T$.  Using this figure,
one can compare the transition temperatures for the LOFF1 and Sarma
phases.  For the latter, $T_c$ is read off as the upper transition
temperature in the 3 plots.  (The lower $T_c$ is where $n_s^{Sarma}(T)$
changes sign and the order parameter vanishes).  It can be seen that the
transition temperatures for the LOFF1 phase are very low compared to
their counterparts in the Sarma phase.

In summary, in this paper we have addressed homogeneous systems and
mapped out the LOFF1 phase diagram at general $T$. We have shown that
LOFF1 phases are more stable near unitarity at sufficiently high $p$
than alternative Sarma states.  Since it is generally expected
\cite{Combescot} that LOFF states with multiple values of $\bf q$ are
more stable than the simplest LOFF1 states, we argue that quite possibly
there exist stable generalized LOFF ground states throughout most (but
perhaps not all) of the dotted regions shown in Figs.~2 and 3; this
corresponds to where the Sarma phase has negative superfluid density.
Our results support previous BdG-based approaches
\cite{Kinnunen,Machida2} which argue that the ground state at unitarity
is a generalized LOFF phase.  In a trap configuration, one might expect
that, since we find the LOFF1 phase is stable at relatively high $p$,
generalized LOFF phases should appear in the neighborhood of the
condensate edge, as also found in earlier work \cite{Kinnunen,Machida2}.
However, in contrast to Ref. \cite{Machida2}, we have addressed
systematic LOFF1 stability criteria and, moreover, find that the size of
the stability region for the LOFF1 state and the size of the existence
region for general LOFF phases very quickly diminish with $T$.
Finally, our self consistent calculations indicate very low $T_c$ in
the LOFF1 state as compared with previous estimates in the literature
\cite{Machida2}, which ignored the effects of non-condensed pairs.

This work was supported by NSF PHY-0555325 and NSF-MRSEC Grant
No.~DMR-0213745.

\bibliographystyle{apsrev}

\end{document}